\documentclass[10 pt, journal]{IEEEtran}
\IEEEoverridecommandlockouts

\usepackage{fancyhdr}

\usepackage[noadjust]{cite}

\usepackage{amsmath}
\usepackage{graphicx}
\usepackage{tabularx}
\usepackage{textcomp}
\usepackage{xcolor}
\usepackage{balance}
\usepackage{comment}
\def\BibTeX{{\rm B\kern-.05em{\sc i\kern-.025em b}\kern-.08em
    T\kern-.1667em\lower.7ex\hbox{E}\kern-.125emX}}
\usepackage[english]{babel}
\usepackage[utf8]{inputenc}
\usepackage{algorithm}
\usepackage[noend]{algpseudocode}
\usepackage{subcaption}
\usepackage[font=small]{caption}
\usepackage{multirow}
\usepackage{graphics}
\usepackage{adjustbox}
\usepackage{soul}
\usepackage{enumitem}
\usepackage{tabularx}
\usepackage{array}
\usepackage{wrapfig}
\usepackage{url}

\usepackage{booktabs}
\usepackage{graphicx}
\begin{document}

\title{SiTe CiM: Signed Ternary Computing-in-Memory for Ultra-Low Precision Deep Neural Networks}

\author{\IEEEauthorblockN{Niharika Thakuria, Akul Malhotra, Sandeep K. Thirumala, Reena Elangovan, Anand Raghunathan and Sumeet~K.~Gupta}\\
\IEEEauthorblockA{
\textit{Purdue University}\\
West Lafayette, Indiana}

}









\maketitle

\pagestyle{plain}

\begin{abstract}
Ternary Deep Neural Networks (DNN) have shown a large potential for highly energy-constrained systems by virtue of their low power operation (due to ultra-low precision) with only a mild degradation in accuracy. To enable an energy-efficient hardware substrate for such systems, we propose a compute-enabled memory design, referred to as SiTe-CiM, which features  computing-in-memory (CiM) of dot products between signed ternary (SiTe) inputs and weights. SiTe CiM is based on cross-coupling of two bit cells to enable CiM of dot products in the signed ternary regime. We explore SiTe CiM with 8T-SRAM, 3T-embedded DRAM (3T-eDRAM) and 3T-ferroelectric metal FET (FEMFET) memories. We propose two flavors of this technique, namely SiTe CiM I/II. In SiTe CiM I, we employ two additional transistors per cell for cross-coupling, achieving fast CiM operations, albeit incurring an area overhead ranging from 18\% to 34\% (compared to standard ternary memories). In SiTe CiM II, four extra transistors are utilized for every 16 cells in a column, thereby incurring only 6\% area cost (but leading to slower CiM than SiTe CiM I). Based on the array analysis, our designs achieve up to 88\% lower CiM latency and 78\% CiM energy savings across various technologies considered, as compared to their respective near-memory computing counterparts. Further, we perform system level analysis by incorporating SiTe CiM I/II arrays in a ternary DNN accelerator and show up to 7X throughput boost and up to 2.5X energy reduction compared to the near-memory ternary DNN accelerators.   
\end{abstract}


\vspace{-0.1in}

\section{Introduction}
\label{sec:introduction}

Deep Neural Networks (DNNs) have revolutionized a variety of fields such as computer vision, natural language processing, robotics, etc. due to their ability to transcend human accuracy for a multitude of complex cognitive tasks \cite{superhuman}. However, this success has come at the cost of enormous computation and storage demands, which has limited their deployment in energy-constrained applications such as edge devices \cite{dl_edge}. DNN compression using reduced precision weights and activations is a widely adopted approach to decrease network complexity and resources \cite{reduced_precision1,reduced_precision2,reduced_precision3,reduced_precision4}. Many works have demonstrated aggressive scaling to ultra-low precisions, all the way up to binary precision (one bit per parameter) \cite{reduced_precision1,reduced_precision2,reduced_precision3,reduced_precision4}. Amongst such ultra-low precision DNNs, ternary precision, in which DNN parameters are quantized to \{-1,0,1\}, has been proposed \cite{ternary_precision1,ternary_precision2,ternary_precision3} as a sweet spot of trade-off for energy and accuracy needs. Offering large energy savings, it suffers a mild loss in accuracy compared to full precision DNNs while showing significant accuracy improvements over binary DNNs, especially for complex tasks. 

An eventual objective of optimizations such as reduced precision at the software level is that they are successfully realized onto a hardware substrate. After commercial success of hardware accelerators with reduced precision (e.g., 4, 8-bit fixed point implementations \cite{tpu}, \cite{nvidea}), ongoing efforts have continued to push the boundary towards designing accelerators that support ultra-low precisions (2-4 bits) \cite{ultra_low1}, \cite{ultra_low2} This motivates us to explore hardware accelerator designs for ternary precision leveraging its amicable trade-offs for energy and accuracy. 

The paradigm of  computing-in-memory (CiM) is a great fit for hardware acceleration since it alleviates concerns faced by von-Neumann based accelerators. Although traditional accelerators such as TPU and GPU Tensor cores \cite{tpu}, \cite{nvidea} leverage their specialized architectures for vector-matrix multiplications (which form the dominant kernels in DNNs), they suffer from the memory wall problem caused by repeated interaction of the processor with memory. CiM performs vector-matrix multiplications within a memory macro, thereby limiting processor-memory calls and mitigating the memory wall problem \cite{imc1}, \cite{imc2}. Over time, CiM has been successfully demonstrated using CMOS based memories such as SRAM \cite{imc1} as well as post-CMOS non-volatile memories (NVMs) such as RRAM, STT MRAM and FEFET \cite{imc2}. While the former offers low write power, maturity of technology and high endurance, NVMs bring attractive attributes for DNNs such non-volatility, high area efficiency and zero stand-by leakage.

Given a large promise of ultra-low precision DNNs and CiM for edge computing, energy efficient hardware accelerators coupling both these design aspects have been in the fore \cite{imc1}, \cite{imc2}. Several of them have been designed for signed binary inputs and signed binary weights, unsigned binary inputs and signed binary weights, signed binary inputs with signed ternary weights and unsigned binary inputs with signed ternary weights. For example, XNOR-SRAM \cite{xnor_sram} and XNOR-RRAM \cite{xnor_rram} have demonstrated CiM for signed binary inputs and weights. Other proposals include 6T-SRAM \cite{xnor_sram} and embedded-DRAM (eDRAM) \cite{dram_cim1} based ternary cells that function with ternary weights and binary operands/activations.  However, little work has been done in the design of CiM-based primitives in the signed ternary regime, with \textit{both} weights and activations encoded as signed ternary. Such designs can be highly useful not only for ternary precison DNNs but potentially can also be generalized for higher precision DNNs with signed activation functions such as transformer models \cite{transformer}. To that end, an in-memory DNN architecture utilizing a specially designed SRAM \cite{tim_dnn} was proposed that performs dot product computation in the signed ternary regime and shows the benefits of massive parallelism, low energy and small accuracy loss. However, this comes at the cost of area. Taking this effort forward, we explored a signed ternary hardware accelerator design with ferroelectric metal transistor (FEMFET) in \cite{tec_cell}. The proposed 3T-FEMFET based ternary cell is realized with an overhead of two additional transistors which are used to cross-couple two underlying FEMFET bit-cells. This resulted in a $\sim$3.3X lower cell area than the SRAM based implementation \cite{tim_dnn}, \cite{tec_cell}. In this paper, we expand on our FEMFET-based design \cite{tec_cell} by exploring two techniques to achieve cross-coupling between memory cells for signed ternary CiM and analyzing their implications for three memory technologies.  The specific contributions of this work are outlined below.
\begin{itemize}
    \item We propose designs of \underline{si}gned \underline{te}rnary CiM (SiTe CiM) for two CMOS memories, (a) 8T-SRAM, and (b) 3T-eDRAM, and (c) a post-CMOS non-volatile 3T-FEMFET memory. We focus on two flavors of SiTe CiM: (a) SiTe CiM I for optimizing latency-energy and (b) SiTe CiM II aimed at higher area efficiency.
    \item We perform array level analysis of SiTe CiM I/II based on 8T-SRAM, 3T-eDRAM and 3T-FEMFET cells and compare their energy, performance and area metrics with the respective near-memory baseline designs. 
    \item We evaluate the system-level performance and energy of SiTe CiM I/II by incorporating the proposed arrays in a ternary DNN accelerator for a wide range of state-of-the-art benchmarks, including both deep convolutional and recurrent neural networks. 
\end{itemize}
\section{Background and Related Work}
\label{sec:background}

\begin{figure}[t!]
\centering
  \includegraphics[width = 0.98\linewidth]{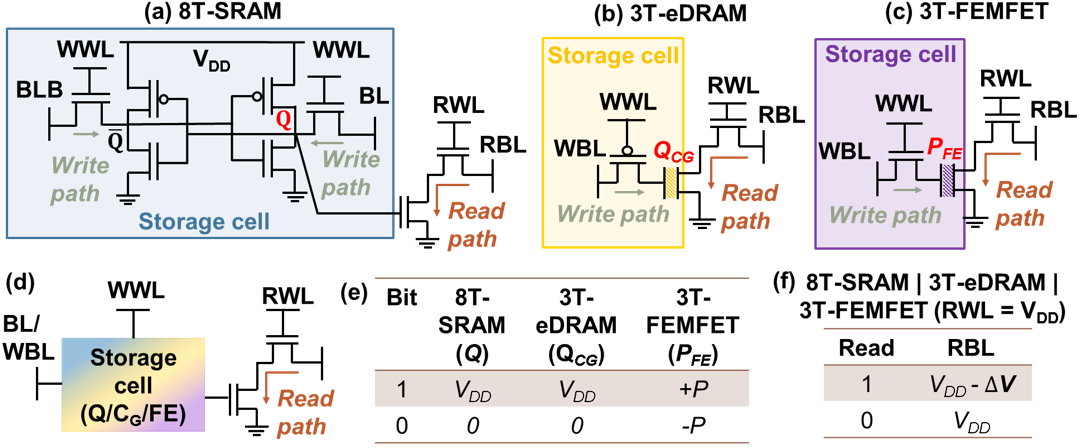}
  \caption{(a) 8T-SRAM, (b) 3T-eDRAM, (c) 3T-FEMFET, (d) Generalized schematic of a bit cell with separated write/read path, (e) bit storage information and (f) bit sense/read information in 8T-SRAM, 3T-eDRAM, 3T-FEMFET.}
  \label{fig:cells}
\end{figure}

In this section, we provide a brief background of the memories (8T-SRAM \cite{8tsram}, 3T-eDRAM \cite{3tdram} and 3T-FEMFET \cite{femfet} based cells - Fig. ~\ref{fig:cells} (a)-(c)) which are utilized for designing SiTe CiM cells in this work. We also provide an overview of the related CiM techniques utilizing these memories. Note, all these memories feature the separation of read and write paths (Fig. ~\ref{fig:cells}(a-d)). This helps in mitigating read-write conflicts for standard memory operation \cite{8tsram}. In the context of CiM, the separation of the read/compute and write paths decouples the effect of the design modifications in SiTe CiM on weight programming, and therefore, allows us to concentrate our discussion on inference. While SiTe-CiM can also be utilized for other memories, including those with shared read and write paths, its impact on the write operation needs a separate analysis. However, later in this paper, we provide a brief qualitative discussion of the implications of SiTe CiM for memories other than those considered in this work, where we present some outlook for the effect on the write operation for memories such as 1T-1R non-volatile memories. For now, we will focus on 8T-SRAM, 3T-DRAM and 3T-FEMFET.

\subsection{8T SRAM}

8T-SRAM (Fig. ~\ref{fig:cells}(a)) employs cross-coupled inverters (with storage nodes $Q$ and $\overline{Q}$) to store the data (or binary weight), and utilizes two write access transistors for programming and a read port for read and CiM \cite{8tsram}. Write or programming of the weights is performed by driving the bitlines $BL$ and $BLB$ to 0/$V_{DD}$ and asserting the write wordline $WWL$. Read is performed by sensing a discharge of voltage ($\Delta V$) from $RBL$ (which is pre-charged to $V_{DD}$) after read wordline $RWL$ is asserted (Fig. ~\ref{fig:cells}(f)). A discharge of $RBL$ voltage ($\Delta V$) from its pre-charged value of $V_{DD}$ indicates a stored value of $V_{DD}$ / ‘1’, while if $RBL$ is maintained at $V_{DD}$, a ‘0’ is sensed. 

For CiM of the dot product between the inputs and stored weights, multiple $RWL$ voltages are applied simultaneously encoding the input vector. Various works have devised CiM primitives for DNNs using the 8T SRAM. \cite{dot_prod} utilizes the 8T SRAM for in-memory multi-bit dot product computation, overcoming the stability issues faced by the 6T SRAM. \cite{twin8t} proposes a CiM primitive for 2-bit weights and activations by employing a twin-8T bitcell. However, the above-mentioned implementations are designed for unsigned inputs targeted towards a subset of neural architectures (such as those with ReLU activation functions). To enable CiM of dot products with signed inputs, either double the number of cycles or double the number of memory arrays would be needed, incurring significant design costs. To mitigate this cost, the work in \cite{tim_dnn} proposes a specialized SRAM design utilizing two 6T SRAMs and 5 control/access transistors per ternary cell to enabling ternary CiM. While the first work of its kind to offer signed ternary CiM, this design suffers from area overheads. In our SiTe CiM designs, we will explore techniques to achieve signed ternary CiM with lower area overheads (quantified later). 

\subsection{3T Embedded DRAM (3T eDRAM)}

 3T-eDRAM (Fig. ~\ref{fig:cells}(b)) is  a popular alternative to conventional 1T-1C DRAMs due to its appealing features such as compatibility with CMOS technology and non-destructive read \cite{3tdram,femfet,1tdram}. In this work, we use an asymmetric 3T-eDRAM (based on \cite{3tdram}) that comprises of an n-type FET whose gate capacitance ($C_G$) serves to store the charge, a p-type write access transistor ($WAX$)  and an n-type read access transistor ($RAX$).  
 
For the write operation, the write  bit line ($WBL$) is driven to $V_{DD}$/0 and $WWL$ is driven to $0$, which charges/discharges $C_G$ to $V_{DD}$/$0$  through $WAX$. $V_{DD}$/$0$ at $C_G$ implies that a value of $1/0$ is stored (Fig. ~\ref{fig:cells}(e)). Its read path  (consisting of the storage FET and $RAX$) and read mechanism is similar to 8T-SRAM, wherein stored information is sensed based on the voltage drop on read bit line ($RBL$) (Fig. \ref{fig:cells}(f)). 

Similar to 8T-SRAM, the CiM of the dot product between the inputs and stored weights can be performed by applying multiple inputs simultaneously as voltages on $RWL$ of each row. CiM designs for DNN acceleration using DRAMs have been explored previously. \cite{dram_cim2} performs a variety of CiM operations using off-the-shelf DRAMs. \cite{dram_cim1} utilizes a 4T DRAM-based bitcell for in-memory ternary-weight binary-input DNN acceleration. Similar to most of the previous 8T SRAM CiM approaches,  these DRAM-based implementations are designed for unsigned inputs. This work aims to address this limitation by achieving in-situ computation of dot product of signed ternary inputs and weights in the eDRAM arrays.

\subsection{FEMFET based Non-volatile Memory (NVM)}

The post-CMOS era of memory technology has observed a growing pool of non-volatile memories (NVM) \cite{imc2}. NVMs exhibit several attractive features such as high area efficiency and zero stand-by leakage. In this work, we choose hafnium zirconium oxide (HZO)-based ferroelectric metal transistor (FEMFET) NVM  \cite{femfet}. Apart from the standard attributes of NVMs discussed in Section ~\ref{sec:introduction}, FEMFETs offer low power electric field-driven write and excellent CMOS compatibility. Amongst the ferroelectric based technologies, FEMFETs offer (i) reduced trap-induced endurance/retention degradation and lower write voltage compared to ferroelectric FETs or FEFETs \cite{femfet}; (ii) significantly larger distinguishability between memory states and decoupled read and write paths compared to FERAMs. However, these benefits come at the cost of area and some gate leakage-induced read challenges \cite{sandeepGL}.    

FEMFETs store bit information in the form of polarization ($P$) of FE in a non-volatile fashion. $+P/-P$ correspond to a low/high resistance state ($LRS/HRS$) and encode bits ‘1’/’0’ (Fig. ~\ref{fig:cells}(c)). While FEMFETs can be utilized to store multi-level values, here, we focus on binary FEMFET devices. We utilize the 3T-FEMFET-based cell (Fig. ~\ref{fig:cells}(c)) consisting of n-type read and write access transistors \cite{tec_cell} connected to the drain and gate of FEMFET respectively (Fig. ~\ref{fig:cells}(c)). For write, we first perform a global reset on all the FEMFET cells (switch polarization to $-P$) by driving the write bit line $WBL$ to -$V_{Write}$. We then selectively set (write $+P$) in some FEMFET cells by driving the corresponding $WBL$ and $WWLs$ to $V_{Write}$.  During read, $RBL$ is pre-charged to $V_{DD}$ and the read word line ($RWL$) is asserted. If the FEMFET stores a ‘1’ (Fig. ~\ref{fig:cells}(f)), RBL discharges, else it remains at $V_{DD}$. FEMFET read can also be performed using current-based sensing. Following the works in \cite{suman_femfet}, we assume that the effect of the gate leakage on read \cite{sandeepGL} can be minimized. 

The study of FEFET/FEMFET-based CiM macros is a promising area of research. \cite{fefet_cim1} proposes CiM-based binarized neural networks using FEFETs. \cite{fefet_cim2} uses FEFETs to implement a ternary content-addressable memory (TCAM) for one-shot learning.  Recently, we explored signed ternary hardware accelerator design by cross-coupling two 3T-FEMFET bitcells \cite{tec_cell}, which achieved signed-ternary CiM, but incurs some area overhead. In this work, we expand on this design by exploring how the benefits of cross-coupling can be attained for signed ternary cells while incurring lower area overheads.  

With this background, we briefly explain our modeling and simulation approach next.

\subsection{Modeling and Simulation}

We design our 8T-SRAM, 3T-eDRAM and 3T-FEMFET based SiTe-CiM cells with minimum sized FETs and FEMFETs for high integration density and low array capacitance. Note, we design the FEMFET with the same cross-section area of the FE and the underlying transistor, which allows us to use minimum size for the underlying transistor.  We use 45nm Predictive Technology Model (PTM) for FETs in 8T-SRAM and 3T-eDRAM \cite{ptm}. For 3T FEMFET cell, 45nm PTM models are used for the access transistor as well as the underlying transistor of FEMFET.  For the ferroelectric, we utilize our modeling framework based on Preisach-based Miller's equation from our earlier work \cite{atanu_drc}. The framework couples the FE self-consistently with PTM model of the underlying FET. We extract the parameters used in Miller's model by calibrating to experimental results \cite{exp_fe}: remanent polarization ($P_R$) = 27 $\mu C/cm^2$, saturation polarization ($P_S$) = 30 $\mu C/cm^2$ and coercive field ($E_C$)= 2.3 MV/cm. We use polarization switching time constant = 200 ps and FE with thickness (T\textsubscript{FE}) = 15 nm. For 8T-SRAM and 3T-eDRAM, the programming voltage is 1V, while for 3T-FEMFET, we use -5V for reset and 4.8V for set.  The supply voltage for CiM and read for all designs presented here is 1V.

\section{Signed Ternary (SiTe) Computation-in-Memory (CiM) Arrays Utilizing Cross-Coupling of Bitcells: SiTe CiM I}
\label{sec:design}

\begin{figure}[t!]
\centering
  \includegraphics[width = 0.95\linewidth]{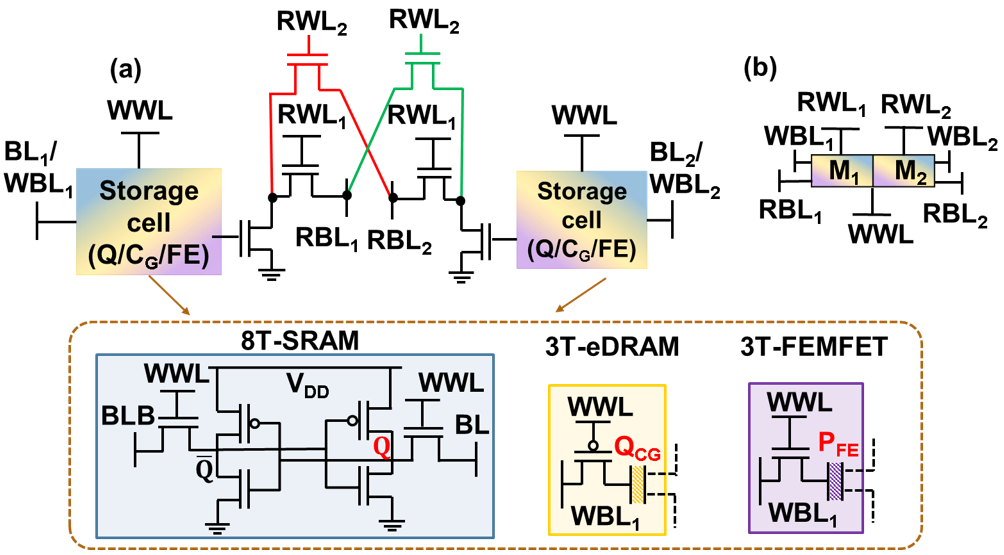}
  \caption{(a) Schematic of SiTe CiM I cell. Inset shows that the underlying cell/storage element can be 8T-SRAM, 3T-eDRAM or 3T-FEMFET cell, (b) symbol of a SiTe CiM I cell.}
  \label{fig:site_cells}
\end{figure}

In this section, we introduce the first flavor of SiTe CiM design, which we refer to as SiTe CiM I. This design is based on cross-coupling two bitcells (storing the ternary weight) to enable dot product computation of signed inputs and signed weights  i.e. $inputs \in [-1, 0, 1]$ and $weights \in [-1, 0, 1]$.  Cross-coupling between bitcells is achieved via two additional access transistors.  We incur an overhead of two transistors per ternary cell in this flavor (Fig. ~\ref{fig:site_cells}(a)). However, this overhead can be reduced with the second flavor of SiTe CiM that we present in the next section (albeit at the cost of CiM performance). Compared to previous implementations of the ternary cells \cite{tim_dnn}, SiTe CiM I incurs smaller area cost, as discussed subsequently.

\renewcommand\thesubsectiondis{\arabic{subsection}.} 
\renewcommand\thesubsubsectiondis{\alph{subsubsection}.}
\renewcommand\theparagraphdis{\roman{paragraph}.}

\subsection{SiTe CiM I Cell}

SiTe CiM I cell consists of two memory bitcells ($M_1$, $M_2$) at its core (Fig. ~\ref{fig:site_cells}). In this work, $M_1$ and $M_2$ correspond to 8T-SRAMs, 3T-eDRAMs or 3T-FEMFETs (Fig. ~\ref{fig:site_cells}(a) and inset). $M_1$ and $M_2$ store binary bit information ‘0’ (high resistance state (HRS)) or ‘1’ (low resistance state (LRS)) either in the form of (a) charge (SRAM/eDRAM), or (b) polarization (FEMFET). Combination of stored weights in $M_1$/$M_2$ constitute a ternary weight ($W$) (Fig. 3(a)), for which we employ differential encoding. For $W$ = 0, both $M_1$ and $M_2$ = 0 (HRS). For $W$ = 1(-1), $M_1$ = 1(0) and $M_2$ = 0(1). SiTe CiM I cell (Fig. ~\ref{fig:site_cells}) features a write wordline ($WWL$) associated with write access transistor(s) and bit lines (e.g., $BL_{1,2}$/$BLB_{1,2}$ in 8T-SRAM, $BL_1$/$BL_2$ in 3T-eDRAM and 3T-FEMFET) for programming $M_1$ and $M_2$. Each of $M_1$ and $M_2$ possess read access transistors, $AX_1$ and $AX_2$ connected to read bit lines $RBL_1$ and $RBL_2$, respectively and controlled with the read word line ($RWL_1$). Further, we add access transistors, $AX_3$ and $AX_4$ and a second read wordline ($RWL_2$) in SiTe CiM I to cross-couple $M_1$ and $M_2$ for signed ternary CiM (Fig. ~\ref{fig:site_cells}(a)). Notice that $AX_3$ connects $M_1$ to $RBL_2$ while $AX_4$ connects $M_2$ to $RBL_1$ in a cross-coupled fashion when $RWL_2$ is asserted. Only $RWL_1$ is asserted for read, while $RWL_1$ or $RWL_2$ are asserted for compute in accordance with ternary input ($I$) encoding (Fig. 3(b)). Like $W$, differential encoding is used for $I$. When $I$ is 0, both $RWL_1$ and $RWL_2$ are 0. When $I$ is 1(-1), $RWL_1$ is driven to $V_{DD}$ (0) and $RWL_2$ is driven to 0 ($V_{DD}$). Interaction of ternary weight ($W$) and input ($I$) produces ternary scalar product ($O$), that is encoded based on combination of $RBL_1$ and $RBL_2$ voltages (Fig. 3(c) - discussed later). For voltage-based sensing, when both $RBL_1$ and $RBL_2$ remain precharged at $V_{DD}$, $O$ is 0. When $RBL_1$ discharges to $V_{DD} - \Delta V$ and $RBL_2$  remains at $V_{DD}$, $O$ is 1. Similarly, when $RBL_2$ discharges to $V_{DD} - \Delta V$ and $RBL_1$  remains at $V_{DD}$, $O$ is -1. In a similar way, we can use current-based sensing, in which current through RBLs encodes the scalar product. For SiTe-CiM 1, we employ voltage-sensing due to its higher energy efficiency compared to current-sensing. Next, we discuss write, read, and CiM operations in SiTe CiM I.

\begin{figure}[t!]
\centering
  \includegraphics[width = 0.98\linewidth]{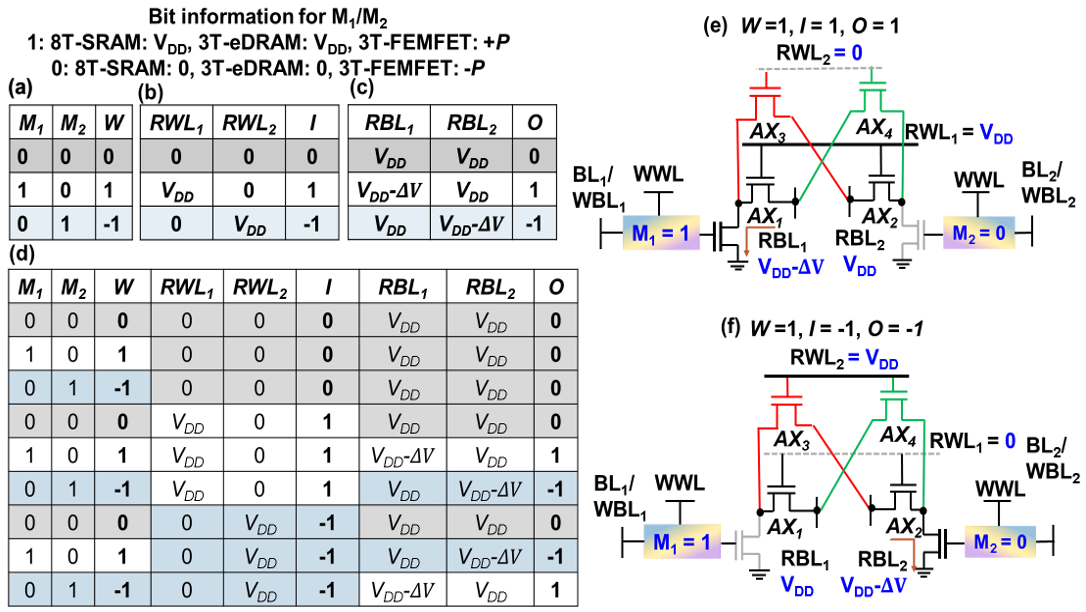}
  \caption{(a) Weight, (b) input, (c) output encoding, (d) truth table for ternary scalar product computation, (e-f) examples of scalar multiplication in SiTe CiM.}
  \label{fig:site_encoding}
\end{figure}

\subsubsection{Write and Read Operations in SiTe CiM I}

\paragraph{Write}
Writing into a SiTe CiM I cell is equivalent to writing into the two binary memory elements, $M_1$ and $M_2$ (as described in Section ~\ref{sec:background}). Based on the ternary value ($W)$ to be written into the cell, $M_1$ and $M_2$ can be programmed according to the encoding in Fig. 3(a). 

\paragraph{Read}
Read is performed by first pre-charging $RBL_1$ and $RBL_2$ to $V_{DD}$ and asserting $RWL_1$ to $V_{DD}$. When sensing $W$ = 1 (-1) , $RBL_2$ ($RBL_1$) remains at $V_{DD}$ while $RBL_1$ ($RBL_2$) drops to $V_{DD} - \Delta V$ (Fig. 3(c)). For $W$ = 0, both $RBL1$ and $RBL2$ remain at $V_{DD}$. Note that during read, $BL$/$BLB$ in 8T-SRAM are kept in a pre-charged state ($V_{DD}$)  whereas for 3T-eDRAM/FEMFET, $WBL$ is kept at 0V. It should be noted that $RWL_2$, which enables the cross-coupling between $M_1$ and $M_2$, is kept off during read and is only used for the signed ternary CiM operation. Thus, the read operation is equivalent to performing conventional reads on $M_1$ and $M_2$ in parallel. 

\subsubsection{Scalar Product Computation in SiTe CiM I}
Ternary scalar product ($O = I*W$) is obtained as a combined outcome of $RBL_1$ and $RBL_2$ voltages (Fig. ~\ref{fig:site_encoding}(c-d)) in response to different ternary weight ($W$) and input ($I$) combinations. To initiate the scalar product computation, both $RBL_1$ and $RBL_2$ are precharged to $V_{DD}$. Next, $RWL_1$ and $RWL_2$ are asserted based on the input value (Fig. ~\ref{fig:site_encoding}(b)). Let us consider different cases on-by-one below:

\paragraph{$I$ = 1}
For this input, $RWL_1 = V_{DD}$, $RWL_2 =$ 0 and $O = I*W = W$. Thus, the scalar product in this case is identical to reading out the weight value. For $W$ = 0, both $RBL1$ and $RBL2$ remain at $V_{DD}$, hence $O$ = 0 (Fig. ~\ref{fig:site_encoding}(c)). When $W$ = 1 (-1), $O$ = 1 (-1) since $RBL_1$ ($RBL_2$) has a low resistance path to ground through $AX_1$ ($AX_2$), discharging it to $V_{DD} - \Delta V$, whereas $RBL_2$ ($RBL_1$) stays at $V_{DD}$. An example illustrating this case is given in Fig. ~\ref{fig:site_encoding} (e). 

\paragraph{$I$ = -1}
For this input, $RWL_1$ = 0, $RWL_2 = V_{DD}$ and $O = I*W = -W$. Thus, the scalar product in this case is \textit{negative} of the weight value. By asserting $RWL_2$, the cross-coupled transistors $AX_3$ and $AX_4$ are turned on. Consequently, it is the cross-coupling that enables the change of sign of the sensed weight value. When $W$ = 1 (i.e. $M_1=1$ and $M_2=0$), $O$ = -1. This is because $RBL_2$ has a low resistance path to ground through $AX_3$ and $M_1$, discharging it to $V_{DD} - \Delta V$. On the other hand, $RBL_1$ stays at $V_{DD}$ as it is connected to the $M_2$ via $AX_4$ (Fig. ~\ref{fig:site_encoding}(c)). Similarly, when $W$ = -1 (i.e. $M_1=0$ and $M_2=1$), the opposite happens (i.e. $RBL_1$ discharges while $RBL_2$ stays at $V_{DD}$ due to cross-coupling), producing $O=1$. When $W$ = 0, both $RBL_1$ and $RBL_2$ remain at $V_{DD}$ leading to $O$ = 0. An example illustrating this case is given in Fig. ~\ref{fig:site_encoding} (f).  

\paragraph{$I$ = 0}
For this input, both $RWL_1$ and $RWL_2$ = 0, and $O = I*W = 0$. Since both read wordlines are off, all the read access transistors ($AX_{1,2,3,4}$) are in off state. Thus, both $RBL_1$ and $RBL_2$ remain at their precharged voltage value of $V_{DD}$. 

\subsection{SiTe CiM I Array and Signed Ternary MAC Operation}
\begin{figure}[t!]
\centering
  \includegraphics[width = 0.98\linewidth]{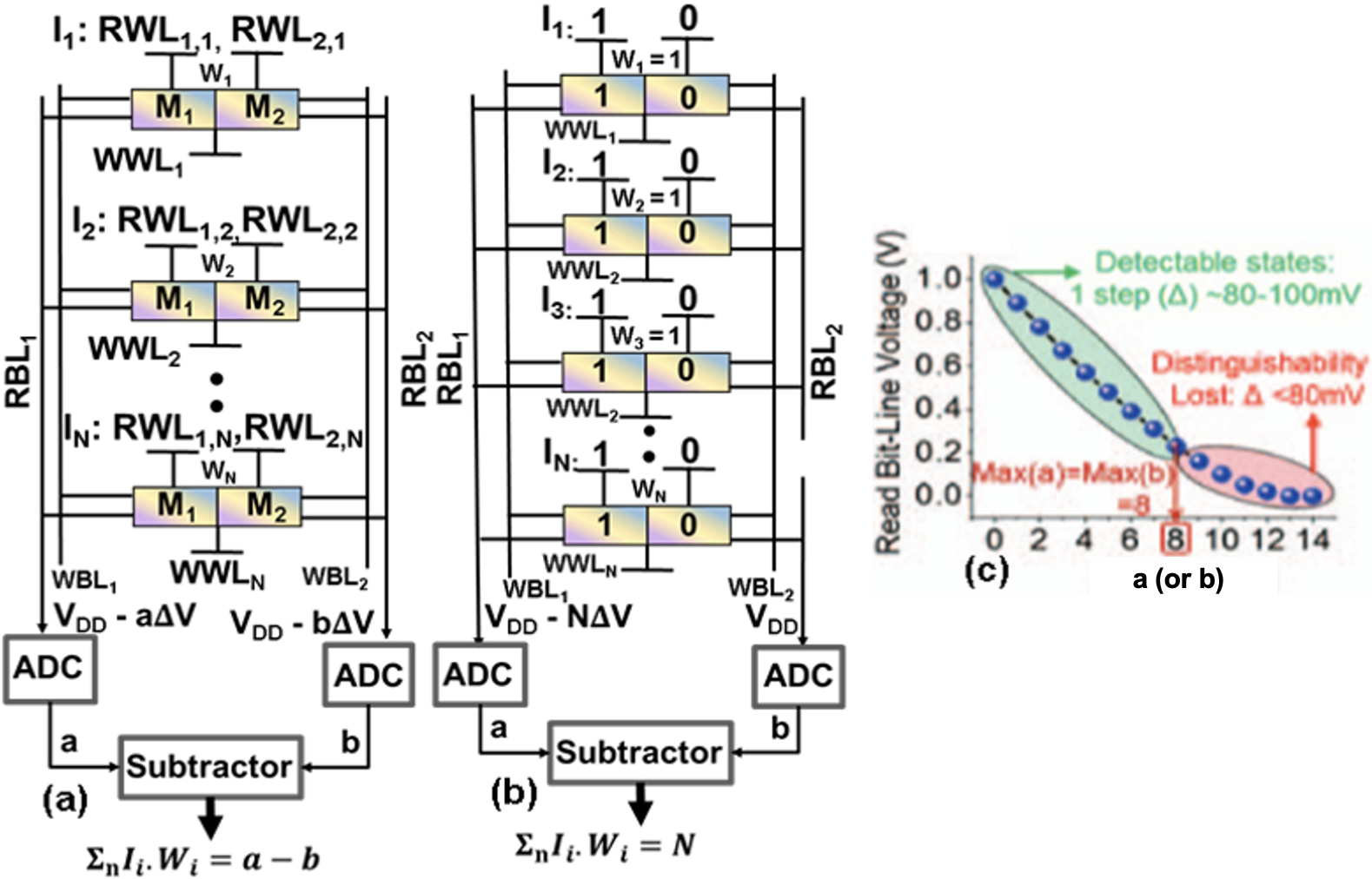}
  \caption{Column of SiTe CiM I cells showing ternary dot-product computation using input ($I$) and weight ($W$) vector, (b) an example of worst-case input-weight combination for maximum $RBL$ discharge, (c) $RBL$ voltage vs number of no. of discharges.}
  \label{fig:site_variation}
\end{figure}

We design 256x256 arrays of SiTe CiM 1 cells (i.e. number of columns \textit{$N_C$} = 256 and number of rows \textit{$N_R$} =256). To accomplish signed ternary MAC computation, we assert multiple rows of the memory array simultaneously in accordance with their respective input values.  We then sense the $RBL_1$ and $RBL_2$ voltages resulting from the collective action of the proposed ternary cells corresponding to the accessed rows. Recall that $RBL_1$ discharging by $\Delta V$ corresponds to the scalar product output of 1 while $RBL_2$ discharging by $\Delta V$ indicates -1 respectively. Hence, if '\textit{$N_A$}' rows are activated simultaneously, out of which ‘\textit{a}’ scalar multiplications produce an output of 1 and ‘\textit{b}’ scalar multiplications produce an output of -1, then the final $RBL_1$ and $RBL_2$ voltages are $V_{DD} - a\Delta V$ and $V_{DD} - b\Delta V$ respectively \cite{tim_dnn}. Similar to \cite{tim_dnn}, we employ 3-bit flash analog to digital converters (ADCs) on each RBL to yield digital values corresponding to ‘\textit{a}’ and ‘\textit{b}’. (The reasons for choosing 3-bit ADC is discussed subsequently.) The final dot product $\sum_{i=1}^{N} I_{i}*W_{i}$ is given by $a - b$, which is achieved by subtracting ‘b’ from ‘a’ using a 3-bit digital CMOS subtractor \cite{tim_dnn}. Fig. ~\ref{fig:site_variation}(a) illustrates this operation. 

An important design decision is about the number of rows that should be activated simultaneously. While $N_A$=$N_R$ (i.e. activating all the rows in a single cycle) would result in maximum parallelism, it would also lead to the range of  ‘\textit{a}’ and ‘\textit{b}’ to be from 0 to $N_R$ (=256). This has two design concerns. First, sensing 257 levels with enough sense margin is a challenge due to device-circuit non-linearities and limited RBL voltage swing (=1V). Second, high precision ADCs would be needed, which have been shown to incur enormous energy and area costs \cite{adc_less}. Keeping in mind that the ternary designs are primarily targeted for energy-constrained systems, we limit the ADC precision to 3-bits to manage the ADC overheads, following the design in \cite{tim_dnn}. This also helps in managing the non-linear mapping between RBL voltage and the output (\textit{a} or \textit{b}). 

To illustrate this point, we plot the RBL voltage versus the expected ADC output in  Fig. ~\ref{fig:site_variation}(c) and extract the sense margin (SM = half the difference of RBL voltages between two adjacent output values). The results in  Fig. ~\ref{fig:site_variation}(c) are for FEMFET-based SiTe CiM I cells; however, since SRAMs and eDRAM have similar read ports, the trends are similar. It can be observed that when the expected $RBL$ output increases from 1 to 8, SM reduces from 50mV to 40mV and becomes even lower for higher values. We attribute this to the exponential behavior of bit-line capacitance discharging. Therefore, asserting 8 rows in a single cycle provides reasonably large sense margin (~40mV). Hence, the 3-bit flash ADC can distinguish between outputs from 0 to 7 with high robustness. We employ an extra sense amplifier to sense the output value of 8. This has also been shown in our study of the effect of threshold voltage ($V_{TH}$) variation on SM in \cite{tim_dnn} and \cite{tec_cell}. It is worth mentioning that several other CiM designs (such as \cite{mit_cim}) show high robustness with SM $\sim$ 30mV. In our work, we perform a more conservative analysis by constraining SM $>$ 40mV for robust computations.

Although SM reduction with increase in the output imparts cautionary conclusions for outputs $>$ 8, sparsity in DNNs help alleviate this constraint. Specifically, sparsity in DNNs (presence of significant weights/inputs as zero values) lowers probability of occurrence of large output values \cite{tfix}. This directly affects the probability of error in the dot-product which is equal to the product of sensing error probability (associated with sense margins) and the occurrence probability of an output value. Our works in \cite{tim_dnn} and \cite{tec_cell} show that although SM for outputs $>$ 8 is lower than the target value of 40mV, it is possible to activate up to 16 rows in a single cycle, leveraging sparsity and low occurrence probability of high-range outputs. With this, the total probability of error has been obtained to be equal to 3.10e-3 \cite{tec_cell}, which the system-level evaluations have shown to have a negligible impact on inference accuracy \cite{tim_dnn}. This is attributed to the inherent resiliency of DNNs to computational errors. Since the target workloads used in these works and our analysis are the same, we follow the design guidelines from \cite{tim_dnn}, Specifically, we assert 16 shows in a single cycle. This yields the maximum output in a cycle to be 8 and all outputs between 8 and 16 are approximated to be equal to 8.

\section{Signed Ternary (SiTe) Computation-in-Memory (CiM) Arrays Utilizing Cross-Coupling of Sub-Columns: SiTe CiM II}
\label{sec:design2}

\begin{figure*}[t!]
\centering
  \includegraphics[width = 0.97\textwidth]{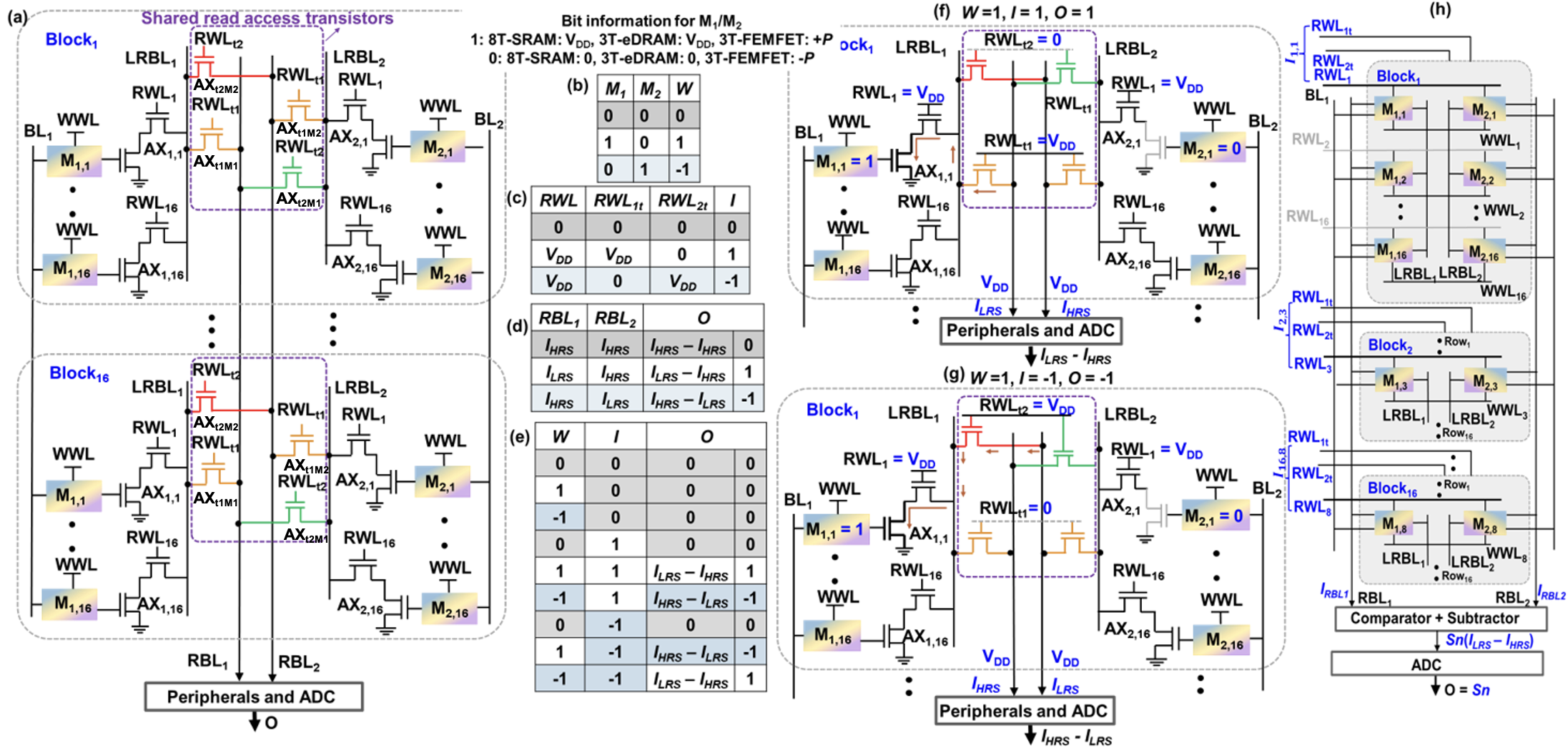}
  \captionsetup{justification=centering}
  \caption{(a) Schematic of a column of SiTe CiM II cells, (b) weight, (c) input, (d) output encoding, (e) truth table for ternary scalar product computation, (f-g) examples of scalar multiplication in SiTe CiM II, (h) dot product computation in a column of SiTe CiM II cells.}
  \label{fig:site2}
\end{figure*}

Let us now present the second flavor of the SiTe CiM design (referred to as SiTe CiM II), which aims to mitigate the area overheads of SiTe CiM I. Instead of employing two cross-coupled transistors ($AX_3$ and $AX_4$) per ternary cell (as in SiTe CiM I), the SiTe CiM II design utilizes four additional transistors for every \textit{16} ternary cells of a column (a sub-column)  to facilitate ternary CiM by means of cross-coupling  (Fig. ~\ref{fig:site2}(a)). Based on this arrangement, we define a block to comprise 16 (rows) x 256 (columns). An array is then formed with 16 blocks or 256 (rows) x 256 (columns).  Due to the sharing of the cross-coupling transistors amongst 16 ternary cells,  we achieve significant area benefits over SiTe CiM I, which is quantified in Section ~\ref{sec:results}. 

In SiTe CiM II (Fig. ~\ref{fig:site2}(a)), the storage elements ($M_1$ and $M_2$) and write mechanisms (using $BL$s/$WBL$s and $WWL$) are similar to SiTe CiM I (Fig. 2(a)). The difference between SiTe CiM I and SiTe CiM II is in the design of their read/compute paths. Notice that in SiTe CiM II, read access transistors $AX_1$ and $AX_2$ associated with $M_1$ and $M_2$ connect to local read bit lines, $LRBL_1$ and $LRBL_2$ respectively (Fig. ~\ref{fig:site2}(a)). $LRBL_1$ and $LRBL_2$ are shared by the 16 cells in a sub-column. In contrast, $RBL_1$ and $RBL_2$ are shared by all (256) cells in the entire column. $AX_1$ and $AX_2$ are activated by $RWL$ that runs along the row (256 cells). In addition, four other access transistors $AX_{t1M1}$, $AX_{t1M2}$, $AX_{t2M1}$ and $AX_{t2M2}$ (marked as shared transistors in Fig. ~\ref{fig:site2}(a)) are used in SiTe CiM II. $AX_{t1M1}$ and $AX_{t1M2}$ are accessible through read wordline $RWL_{t1}$. They connect $LRBL_{1}$ and $LRBL_{2}$ to $RBL_1$ and $RBL_2$ respectively. $AX_{t2M1}$ connects $LRBL_1$ to $RBL_2$ and $AX_{t2M2}$ connects $LRBL_2$ to $RBL_1$ using read wordline $RWL_{t2}$. $RWL_{t1}$ and $RWL_{t2}$ are shared along a row. The inclusion of $AX_{t2M1/M2}$ along with $RWL_{t2}$ allows us to encode signed ternary $I$ = -1 which, in turn, realizes cross-coupling of $M_1$ to $RBL_2$ and $M_2$ to $RBL_1$ for signed ternary output (details discussed later). It is important to note that current-based sensing is used in this design instead of voltage-based sensing (which was employed for SiTe CiM I). This is because voltage-sensing may lead to incorrect sensing of binary value ‘0’ caused by charge sharing between $LRBL$ and $RBL$. The said issue further aggravates when multiple rows are activated to accomplish CiM. Current-based sensing helps in circumventing this concern. In other words, while SiTe CiM II poses a restriction for the sensing to be current-based,  SiTe CiM I cells are compatible with both voltage and current-based sensing. Next, we explain the write, read and CiM operations.

\subsection{Write and Read Operations in SiTe CiM II}
\paragraph{Write}
 Similar procedure as SiTe CiM I (discussed previously) is followed to store $W$ in SiTe CiM II cells (Fig. ~\ref{fig:site2}(b)).

\paragraph{Read}
In order to read, $RBL_1$ and $RBL_2$ are driven to $V_{DD}$. Then, $RWL_i$ and $RWL_{t1}$ are asserted while $RWL_{t2}$ is kept at 0V (subscript $i$ denotes the $i^{th}$ row in the block being read). The said biasing creates a path for current to flow from $RBL_1$ ($RBL_2$) via $AX_1$ ($AX_2$) and $AX_{t1M1}$ ($AX_{t1M2}$) when $M_1$($M_2$) stores 1. The current is sensed on $RBL_1$ and $RBL_2$ to determine the stored ternary weight ($W$). For ternary '0' ($M_1$ = 0 and $M_2$ = 0), $I_{HRS}$ is sensed on both $RBL_1$ and $RBL_2$. Note that $I_{HRS}$ here mainly results from a small current that flows from $RBL_2$ to charge the capacitor of $LRBL_2$. Negligible current flows through $M_1$/$M_2$ when either is in HRS.  When sensing ternary ‘1’ ($M_1$ = 1 and $M_2$ = 0), we obtain $I_{LRS}$ and $I_{HRS}$ on $RBL_1$ and $RBL_2$ respectively (similar to Fig. ~\ref{fig:site2}(f)).  For ternary ‘-1’ ($M_1$ = 0 and $M_2$ = 1), currents available on $RBL_1$ and $RBL_2$ are $I_{HRS}$ and $I_{LRS}$, respectively. 

\subsection{Obtaining Scalar Product in SiTe CiM II}
As mentioned earlier, weights are encoded as a combination of bits stored in $M_1$ and $M_2$ like in SiTe CiM I (Fig. ~\ref{fig:site2}(b)). The inputs are encoded as a combination of $RWL$, $RWL_{t1}$ and $RWL_{t2}$ as shown in Fig. ~\ref{fig:site2}(c). The ternary scalar product outputs depend on currents appearing on $RBL_1$/$RBL_2$ ($I_{RBL1}$/$I_{RBL2}$) and is equal to their subtracted value (Fig. ~\ref{fig:site2}(d)). We detail this with examples below: 

\paragraph{$I$ = 1}
For this input, $RWL$ = $V_{DD}$, $RWL_{t1}$ = $V_{DD}$, $RWL_{t2}$ = 0 and $O = I*W = W$. This case is similar to the read operation. Hence, for $W$ = 1 ($M_1$ = 1/LRS and $M_2$ = 0/HRS), we receive $I_{RBL1}$ = $I_{LRS}$ and $I_{RBL2}$ = $I_{HRS}$. As described above, we perform subtraction of the RBL currents to obtain $O$ = $I_{LRS}$ – $I_{HRS}$ = 1 (Fig. ~\ref{fig:site2}(f)). When $W$ = -1 ($M_1$ = 0/HRS and $M_2$ = 1/LRS), $I_{RBL1}$ = $I_{HRS}$ and $I_{RBL2}$ = $I_{LRS}$; $O$ = $I_{HRS}$ – $I_{LRS}$ = -1. For $W$ = 0, $I_{RBL1}$ = $I_{RBL2}$ = $I_{HRS}$, $O$ = 0.

\paragraph{$I$ = -1}
For this input, $RWL$ = $V_{DD}$, $RWL_{t1}$ = 0, $RWL_{t2}$ = $V_{DD}$ and $O = I*W = -W$. Turning on $RWL_{t2}$ enables the cross-coupling between $M_1$ and $M_2$. For $W$ = 1 ($M_1$ = 1, $M_2$ = 0), we achieve $I_{RBL1}$ = $I_{HRS}$ and $I_{RBL2}$ = $I_{LRS}$. Hence, $O$ = $I_{HRS}$ – $I_{LRS}$ = -1 (Fig. ~\ref{fig:site2}(g)). Conversely, when $W$ = -1, $I_{RBL1}$ = $I_{LRS}$, $I_{RBL2}$ = $I_{HRS}$ and $O$ = $I_{LRS}$ – $I_{HRS}$ = 1. In case of $W$ = 0, $O$ = $I_{HRS}$ – $I_{HRS}$ = 0.

\paragraph{I = 0}
For this input, $RWL$ = $RWL_{t1}$ = $RWL_{t2}$ = 0 and $O = I*W = 0$. Since all read word lines are at 0V, all read access transistors are off, preventing current flow through $RBL_1$ and $RBL_2$. Hence, $O$ = 0.

\begin{figure}[t!]
\centering
  \includegraphics[width = 0.98\linewidth]{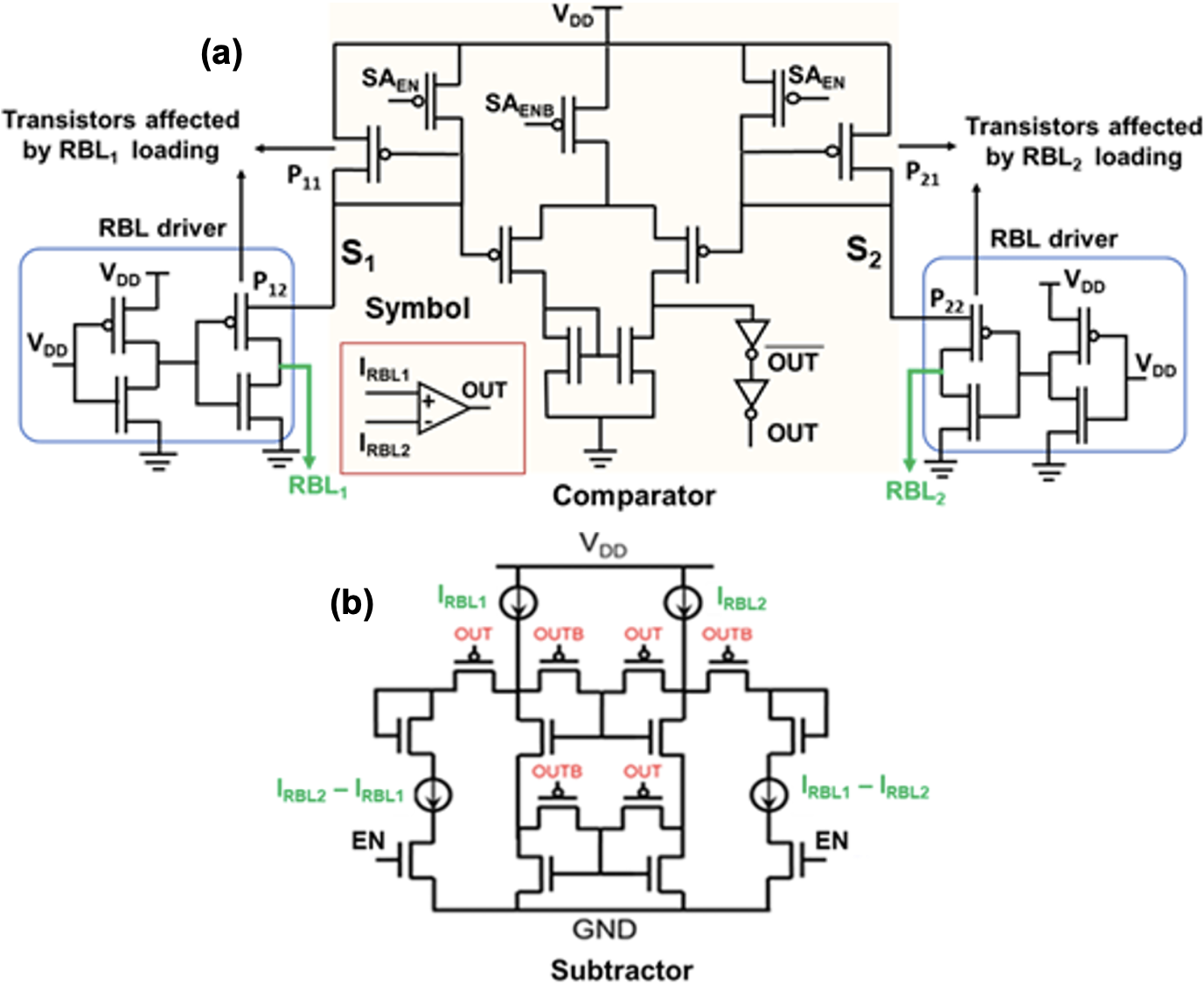}
  \caption{The circuit schematics of the (a) comparator and (b) subtractor, used in SiTe CiM II.}
  \label{fig:comparator}
\end{figure}

\subsection{Signed Ternary MAC Operation using SiTe CiM II}
Like in SiTe CiM I, multiple rows are accessed for computing ternary MAC output. However, multi-row access  in SiTe CiM II is attainable only when each accessed row belongs to \textit{distinct} blocks. We illustrate this in Fig. ~\ref{fig:site2}(h) wherein the accessed row of each block (without any loss of generality) is shown with inputs marked. Recall that read access transistors $AX_{t1M1/M2}$ and $AX_{t2M1/M2}$ in SiTe CiM II are shared within the block, hence a common $RWL_{t1}$/$RWL_{t2}$ is associated with a sub-column or the 16 column cells in a block (Fig. ~\ref{fig:site2}(h)). Thus, for the cells within the same block, distinct values of the input vector must be applied in different cycles. In other words,  one row from each block is asserted in a given cycle to compute MAC output. Hence, it takes \textit{$N_{RB}$} cycles for the entire computation. (where \textit{$N_{RB}$} is the number of rows in a block or number of cells in a sub-column).   

It is noteworthy that even in SiTe CiM I, despite the option to simultaneously apply distinct inputs on all the rows, only 16 rows were accessed in a single cycle. This was done to maintain sufficient compute robustness, sacrificing parallelism to a partial extent. SiTe CiM II takes into account this robustness-driven constraint to group multiple cells in a block and share the cross-coupling transistors amongst those cells. The choice of the size of the block is dictated by the targets for the compute robustness, amount of parallelism, CiM latency (determined in part by the interaction of the LBLs and BLs) and the layout footprint. Here, we utilize the same design choice as SiTe CiM I i.e. \textit{$N_A$}= 16 rows activated in a single cycle. With that, we obtain the number of rows in a block \textit{$N_{RB}$} = \textit{$N_{R}$}/\textit{$N_A$} = 256/16=16. Like SiTe CiM I, the reason for this choice is driven by managing the ADC costs (i.e. restricting its precision to 3-bits) as well as by sustaining the computation robustness, as will be discussed in the next sub-section.

After multiple rows are asserted, currents due to the interaction of $I_i$ and $W_i$ in each row aggregate on $RBL_1$ and $RBL_2$. $I_{RBL1}$ and $I_{RBL2}$ are then used to evaluate the dot-product as follows. First, a comparator is used to determine which of $I_{RBL1}$ and $I_{RBL2}$ has higher current and its output represents the sign of the MAC output ($S$). If $I_{RBL1} > I_{RBL2}$, $S$ = 1, otherwise $S$ = -1. Then the comparator output along with $I_{RBL1}$ and $I_{RBL2}$ is fed to a current subtractor circuit which determines the magnitude
of the difference of $I_{RBL1}$ and $I_{RBL2}$. $|I_{RBL1}-I_{RBL2}|$ is an integral
multiple 'n' of $I_{LRS}$ - $I_{HRS}$. The comparator and subtractor circuit schematics are shown in Fig.~\ref{fig:comparator}. We extract the value of ‘n’ by utilizing an ADC. We provide an example here to illustrate our MAC
computation using SiTe CiM II. Let $I_{RBL1}$ = a$I_{LRS}$+ (16 - a)$I_{HRS}$, $I_{RBL2}$ = c$I_{LRS}$ + (16 - c)$I_{HRS}$ and $I_{RBL2} > I_{RBL1}$. This indicates that the comparator output is $S$ = -1 as per our previous discussion. Then,  we employ the subtractor to get $I_{RBL2}$ - $I_{RBL1}$ = $(c-a)(I_{LRS} - I_{HRS})$. When this is passed to the ADC, we obtain $n = c-a$. The final MAC output is $O = *n = a-c$.

Note, in SiTe CiM I design, we first digitize the RBL output voltage using two 3-bit flash ADCs (one for each RBL) and then subtract the results using a digital subtractor. In SiTe CiM II design, we first subtract the current using the comparator and current subtractor, and then digitize the output using a single 3-bit current-mode flash ADC. While SiTe CiM II saves on 1 ADC (compared to SiTe CiM I), it incurs higher array energy during compute and needs an analog subtractor (as opposed to a digital one). Overall, we find that the voltage-based sensing offers a higher energy efficiency (considering all these trade-offs), so we utilize it for SiTe CiM I design. On the other hand, SiTe CiM II    is restricted to use current-based sensing, as discussed before. However, it provides higher area efficiency, as quantified subsequently. Similar to SiTe CiM I, We employ an extra sense amplifier to sense the output value of 8. 

\subsection{Sense Margin (SM) Analysis of Signed Ternary MAC using SiTe CiM II}

\begin{figure}[t!]
\centering
  \includegraphics[width = 0.98\linewidth]{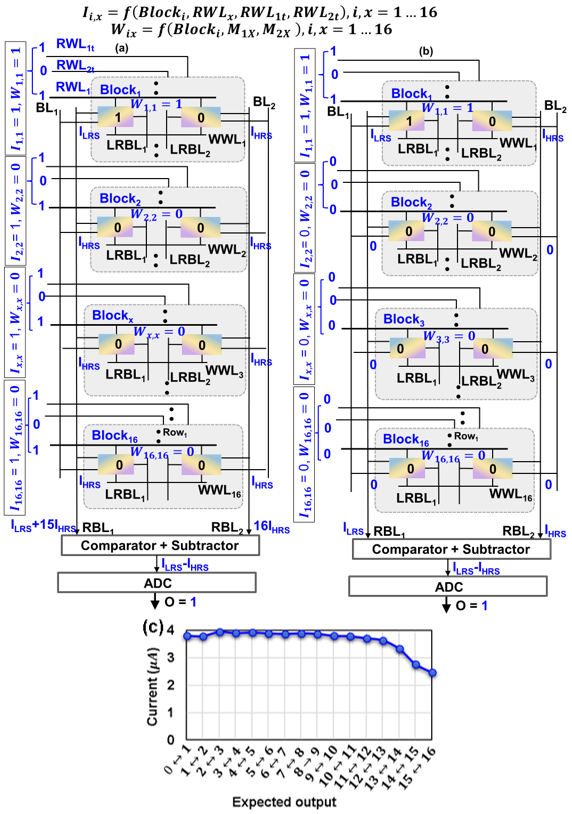}
  \caption{Examples of dot product output = 1 using inputs and weights for (a) maximum loading effect and (b) minimum loading effect, (c) sense margin for expected output 0 through 16.}
  \label{fig:site2_variation}
\end{figure}

Here, we study SM as a function of MAC outputs using SiTe CiM II cells. We perform this analysis to establish the maximum number of simultaneous row assertions for MAC operation similar to the exercise done for the SiTe CiM I cell. Since SiTe CiM II utilized current sensing (as opposed to voltage sensing in SiTe CiM I), SM analysis needs a different treatment, considering the loading effect of the sensing circuitry (which is not needed in voltage sensing with floating $RBL$s). We present our approach of analyzing sense margin for SiTe CiM II using the examples shown in Fig. ~\ref{fig:site2_variation}(a-b). In both cases, we expect $O$ = 1, however the weight and input combinations applied in the two cases are different. In Fig. ~\ref{fig:site2_variation}(a), which we refer to as example (a), we use ($I_1$, $W_1$) = (1,1) whose outcome is $I_{RBL1}$ = $I_{LRS}$ and $I_{RBL2}$ = $I_{HRS}$. For the remaining rows, i.e., row 2-16, we set ($I_{2..16}$, $W_{2..16}$) = (1,0). Corresponding to $W$ = 0 and $I$ = 1 for these rows, we obtain non-zero currents on $RBL_1$ and $RBL_2$ due to their interaction with $L_{RBL1}$ and $L_{RBL2}$, respectively, i.e., $I_{RBL1}$ =15$I_{HRS}$ and $I_{RBL2}$ =15$I_{HRS}$ (Fig. ~\ref{fig:site2}(e)). Combining total current on $RBL_1$ and $RBL_2$ from all the rows, we have $I_{RBL1}$ = $I_{LRS}$ +15$I_{HRS}$ and $I_{RBL2}$ = 16$I_{HRS}$¸ and finally $O$ = $I_{RBL1}$ - $I_{RBL2}$ = $I_{LRS}$ - $I_{HRS}$ = 1.

In Fig. ~\ref{fig:site2_variation}(b), which we refer to as example (b),  we start with ($I_1$, $W_1$) = (1,1), i.e., $I_{RBL1}$ = $I_{LRS}$ and $I_{RBL2}$ =$I_{HRS}$. However, for rows 2-16, our inputs and weights are ($I_{2..16}$, $W_{2..16}$) = (0,0). Since, $I_{2..16}$ = 0, no current contribution on $RBL_1$ and $RBL_2$ is observed from those rows. Overall, we observe $I_{RBL1}$ = $I_{LRS}$ and $I_{RBL2}$ = $I_{HRS}$¸ and $O$ = $I_{RBL1}$ = $I_{LRS}$ - $I_{HRS}$ = 1. 

Notice that for identical outputs in examples (a) and (b), $I_{RBL1/2,a} > I_{RBL1/2,b}$, where $I_{RBL1/2,a}$ and $I_{RBL1/2,b}$ are the read bit line currents for example (a) and (b), respectively. For $O$ = 1, example (a) is the worst case (WC) example leading to the maximum current in the $RBL$s since rows with $W$ = 0 also contribute to $RBL$ current. This results in the largest loading of the sensing circuitry, leading to largest deviation of the output current from the ideal current (i.e. with sensing resistance = 0). Whereas example (b) is the best case (BC) i.e. least loading, since only the row with $W$ = 1 is solely responsible for $RBL$ current resulting in the minimum current. We perform a similar exercise to identify BC/WC examples for all outputs $\leq$ 16 and ultimately obtain the respective $I_{RBL1/2}$ values. Finally, for an expected output $O_{(n-1) \leftrightarrow n}$, we define SM margin as ($O_{BC,n}$ - $O_{WC,n-1}$)/2. Fig. ~\ref{fig:site2_variation}(c) shows SM calculated using this method. We observe from Fig. ~\ref{fig:site2_variation}(c), that the SM begins to diminish for $O$ $>$ 8. This is similar to our prior observation during the analysis of the SiTe CiM I cell, where the SM reduced for $O$ $>$ 8. Thus, SM analysis provides us with a consistent parameter for parallel assertion of rows in both SiTe CiM I and II designs to be used for their array design. That being said, the sparsity present in DNN weights/inputs also plays an important role in determining the degree of parallel row assertion, as discussed in detail in the previous section. Overall, the number of rows to be activated simultaneously is chosen as 16 i.e. \textit{$N_A$} = 16, for SiTe CiM II  accounting for both SM and sparsity, as well as limiting the ADC precision to 3-bits to manage the associated costs (just like SiTe CiM I). This dictates the block size, as discussed previously. Note, like SiTe CiM I. the maximum output in a cycle to 8 and all outputs between 8 and 16 are approximated to be equal to 8.  
\section{Array-Level Analysis of SiTe CiM I/II}
\label{sec:results}

\begin{figure}[t!]
\centering
  \includegraphics[width = 0.98\linewidth]{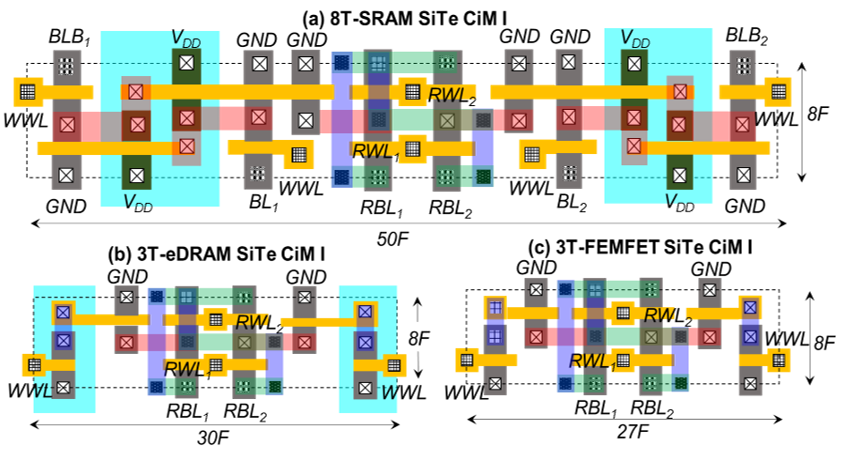}
  \caption{Layout of SiTe CiM I cell based on (a) 8T-SRAM, (b) 3TeDRAM, (c) 3T-FEMFET bit cells.}
  \label{fig:site1_layout}
\end{figure}

Here, we present an array-level analysis of area and write-read-CiM performance and energy of proposed 8T-SRAM, 3T-eDRAM and 3T-FEMFET SiTe CiM I/II arrays by comparing them to their near memory (NM) counterparts. The NM ternary accelerators for the baseline designs read scratchpad memories row-by-row to obtain the weights, computes the scalar products with the corresponding input and accumulates them over multiple cycles to obtain the dot products. The baseline memories are designed with standard 8T-SRAM, 3T-eDRAM and 3T-FEMFET arrays of size 512x256, with a ternary weight stored in two bit-cells (yielding 256x256 ternary cells in total). Thus, two memory cells store one ternary weight. We use voltage-sensing to read the weight values in the baseline designs (as it is more energy efficient than the current-sensing).

\subsection{SiTe CiM I}
\subsubsection{Layout Area}
We show the layout of SiTe CiM I cells designed with 8T-SRAM in Fig. ~\ref{fig:site1_layout}(a), 3T-eDRAM in Fig. ~\ref{fig:site1_layout}(b) and 3T-FEMFET in Fig. ~\ref{fig:site1_layout}(c). 8T-SRAM, 3T-eDRAM and 3T-FEMFET SiTe CiM I cells exhibit 18\%, 34\% and 34\% higher
area, respectively compared to corresponding 8T-SRAM, 3T-eDRAM and 3T-FEMFET NM baseline cells. (Note, the ternary cell in NM baselines comprises 2 bit-cells). The overhead in area is attributed to addition of read access transistors $AX_3$ and $AX_4$ (Fig. ~\ref{fig:site_cells}(a)) that enable ternary in-memory computation in SiTe CiM I. The proposed modification has a larger impact on 3T-eDRAM/FEMFET designs than 8T-SRAM since the baseline cells in the former are smaller than the latter.
We also compare the area of 8T-SRAM based SiTe CiM I cell with the SRAM-based design presented in \cite{tim_dnn}, which was the first design to achieve signed ternary CiM. Our layout analysis shows that our design achieves 44\% lower area compared to the design in \cite{tim_dnn}.  

We also estimate the effect of peripheral circuits such as ADCs in the SiTe CiM I design and multiply-accumulate unit in the NM baseline. Note, ADCs are not required in NM designs as the rows are read sequentially. Similarly, a separate multiply-accumulate unit is not needed in SiTe CiM designs as the dot product computation occurs within the memory array. We observe that the ADCs pose a large area overhead compared to the multiply-accumulate unit. Overall, SiTe-CiM I exhibits 1.3x-1.53x larger area compared to the NM baselines.  

\subsubsection{CiM Energy and Delay}

\begin{figure}[t!]
\centering
  \includegraphics[width = 0.98\linewidth]{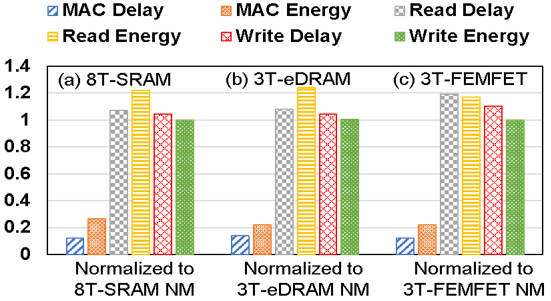}
  \caption{Array level analysis of SiTe CiM I based on (a) 8T SRAM, (b) 3T-eDRAM, (c) 3T-FEMFET. Results are presented after normalization with respective NM counterparts.}
  \label{fig:site1_results}
\end{figure}

SiTe CiM I designs show significant energy (orange bars in Fig. ~\ref{fig:site1_results}(a-c)) and delay improvements (blue bars in Fig. ~\ref{fig:site1_results}(a-c))) for CiM operations across
various technologies considered in this work. This is attributed to massively parallel ternary dot-product computation  in the proposed arrays contrary to row-by-row access in their near-memory counterparts. This results in $\sim$ 88\% lower latency in SiTe CiM I cells for the three technologies compared to the corresponding NM baselines  (blue bars in Fig. 8(a-c)). The benefits of CiM are also reflected in CiM energy of SiTe CiM I cells. 8T-SRAM, 3T-eDRAM and 3T-FEMFET SiTe CiM I cells exhibit 74\% (Fig. ~\ref{fig:site1_results}(a)), 78\% (Fig. ~\ref{fig:site1_results}(b)) and 78\% ((Fig. ~\ref{fig:site1_results}(c))) improved energy efficiency compared to the respective baselines. Since MAC operations
are the predominant contributor of energy in DNNs, we expect energy savings achieved at the array level to be translated to system-level energy efficiency (as discussed in Section ~\ref{sec:system}).

\subsubsection{Read and Write}
SiTe CiM I designs for the technologies considered in this work experience overhead of performance and energy for read/write operations compared to NM baseline. We attribute this to increased cell area and bit line capacitance caused by the additional read access transistors used for the enablement of signed ternary CiM in the proposed designs. 8T-SRAM, 3T-eDRAM and 3T-FEMFET SiTe CiM I cells show 22\%, 24\%, 17\% higher read energy and  7\%, 7\%, 19\% higher read latency, respectively than their NM counterparts (Fig. ~\ref{fig:site1_results}). 

Similarly, our analysis shows comparable write energy and  4\%, 4\%, 10\% degraded write performance (higher latency) in 8T-SRAM,  3T-eDRAM and 3T-FEMFET SiTe CiM I, respectively compared to their baselines (Fig. ~\ref{fig:site1_results}). Note that despite the read-write overheads mentioned above, significant improvement in system performance and energy is observed for ternary DNNs implemented using SiTe CiM I arrays (details in the next section). This is because more than 90\% of operations in DNNs are MAC operations, which overshadows the overheads of the read/write operations.

\subsection{SiTe CiM II}
 
\subsubsection{Layout Area}
\begin{figure}[t!]
\centering
  \includegraphics[width = 0.98\linewidth]{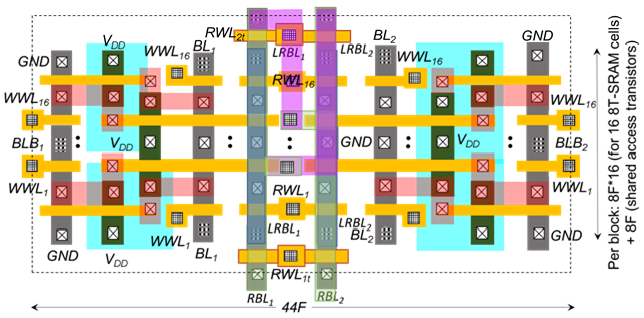}
  \caption{Layout of SRAM-based SiTe CiM II cell}
  \label{fig:site2_layout}
\end{figure}

Recall that we proposed SiTe CiM II with the objective to minimize area overhead caused by additional transistors added in SiTe CiM I for ternary CiM capability. Specifically, we followed the approach of sharing 4 access transistors among 16 cells in a sub-column. Such an approach
allows a layout that can be optimized for 8T-SRAM, 3TeDRAM and 3T-FEMFET based SiTe CiM II 
cells alike. In essence, the shared transistors are laid in a manner that add two poly pitches (8F) to the height of a block (=8F*16). Note that 8T-SRAM, 3T-eDRAM and 3T-FEMFET based cells
have identical cell/block height. Moreover, the cell/block width of SiTe CiM II cells based on 8T-SRAM, 3T-eDRAM and 3T-FEMFET are equal to their respective NM baselines. Considering both these aspects, the overhead in area for all the three technologies is contributed by the increased block height, and is identical for all. We show a layout of 8T-SRAM based SiTe CiM II in Fig.~\ref{fig:site2_layout} to represent this. We observe that the area overhead for SiTe CiM II arrays is 6\%
across all technologies compared to their respective baselines. This can be compared to the area overhead of SiTe CiM I (18\% for 8T SRAM and 34\% and 3T-eDRAM/FEMFET) to appreciate the effectiveness of the proposed sharing of the access transistors in SiTe CiM II.

Including the effect of peripheral circuits (similar to SiTe CiM I), we observe 1.21x-1.33x larger area of SiTe CiM II compared to their NM
counterparts.

\subsubsection{CiM Energy and Delay}
\begin{figure}[t!]
\centering
  \includegraphics[width = 0.98\linewidth]{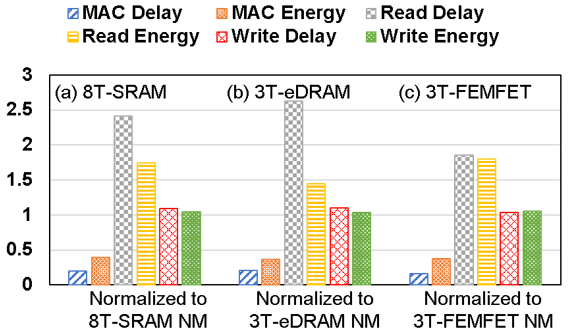}
  \caption{Array level analysis of SiTe CiM II based on (a) 8T SRAM, (b) 3T-eDRAM, (c) 3T-FEMFET. Results are presented after normalization with respective NM counterparts.}
  \label{fig:site2_results}
\end{figure}

Similar to SiTe CiM I, we observe improvements in CiM delay and energy of SiTe CiM II cells over the NM baselines across the three technologies due to massively parallel computation capability of the former. 8T-SRAM (Fig. ~\ref{fig:site2_results}(a)), 3T-eDRAM (Fig. ~\ref{fig:site2_results}(b)), and 3T-FEMFET (Fig. ~\ref{fig:site2_results}(c)) SiTe CiM II cells show 80\%, 78\% and 84\% improvement in MAC delay over their respective NM baselines. At the same time, 8T-SRAM, 3T-eDRAM and 3T-FEMFET SiTe CiM II cells are 61\% (Fig. ~\ref{fig:site2_results}(a)), 63\% (Fig. ~\ref{fig:site2_results}(b)) and 62\% (Fig. ~\ref{fig:site2_results}(c)) more energy efficient than the corresponding NM baselines. 

Note that the benefits in performance and energy reported here are lower compared to SiTe CiM I. This is because of current-based sensing implemented in this design (discussed in Section ~\ref{sec:design}) necessitates that bit-lines are driven from 0 to $V_{DD}$ at the onset of sensing. In contrast, voltage-based sensing (used in the NM baselines and SiTe CiM I design) involves the bitlines being precharged from $V_{DD}$ - x*$\Delta V$ to $V_{DD}$, where x*$\Delta V$ is the discharge caused by the previous CiM operation (discussed in Section ~\ref{sec:design}).  Moreover, other factors such as (i) increased block height leading to higher bit line capacitance, (ii) charging/discharging of an additional read word line present and (iii) lower efficiency of current-based sensing circuitry in SiTe CiM II also add to increased latency and switching energy. As a result, compared to SiTe CiM I, lower performance improvements (over the baselines) are observed in SiTe CiM II. 

\subsubsection{Read and Write}

8T-SRAM, 3T-eDRAM and 3T-FEMFET SiTe CiM II cells show 2.4X, 2.6X and 1.8X lower read speed and 74\%, 44\% and 79\% higher read energy, respectively, compared to the corresponding NM baselines. This is due to (i) higher bit line capacitance due to additional read access transistors and increased block height in SiTe CiM II, (ii) less energy efficient current-based sensing mechanism (discussed previously) and (iii) charging/discharging of an additional read word line. We observe comparable write energy and 8\%, 10\%, 3\% lower write performance in 8T-SRAM, 3T-eDRAM and 3TFEMFET SiTe CiM II cells, respectively compared to their NM counterparts. Similar to SiTe CiM I, we expect significant improvement in system performance and energy for SiTe CiM II based ternary DNNs due to amortization of read/write overhead by the majority CiM operations (which constitute $>$ 90\% of all operations).

\subsection{SiTe CiM I versus SiTe CiM II}

From the results discussed above, it can be concluded that while SiTe CiM I offers higher CiM energy efficiency and performance, SiTe CiM II achieves higher integration density. Specifically, compared to SiTe CiM I, SiTe CiM II exhibits 1.5X, 1.7X and 1.7X lower CiM energy, and 1.7X, 1.8X and 1.3X higher CiM latency for 8T-SRAM, 3T-eDRAM and 3T-FEMFET, respectively. However, SiTe CiM II achieves 10\% lower cell area for 8T-SRAM and 21\% lower cell area for 3T-eDRAM/3T-FEMFET. Thus, depending on the design needs, one can chooses one implementation over the other. 

Next, we evaluate the system-level implications of the two Site CiM designs for the three technologies and compare them to the respective NM baselines    
\section{System Level Analysis}
\label{sec:system}

To evaluate the system-level energy and performance benefits of SiTe CiM I/II designed with 8T-SRAM, 3T-eDRAM and 3TFEMFET cells, we utilize the TiM DNN architecture \cite{tim_dnn}. The partial sums from each column of the SiTe CiM 1/II arrays are stored in peripheral compute unit (PCU) using a sample and hold circuitry and are accumulated after multiple block accesses to result in the
final dot products. We design our macros with 32 PCUs ($<$ $N_C$=256) for the entire array, to limit the overheads due to peripheral circuits \cite{tim_dnn}. Finally, the dot products are quantized, and passed through an activation function to obtain the inputs to the next DNN layer. We perform our analysis with 5 DNN benchmarks, viz. AlexNet, ResNet34, Inception, LSTM and GRU.

\renewcommand\thesubsectiondis{\Alph{subsection}.}

\subsection{Simulation Framework}

We design our iM architecture based on TiM-DNN \cite{tim_dnn} with 32
SiTe-CiM I (SiTe-CiM II) arrays, where each array consists of 256x256 SiTe-CiM I (SiTe-CiM II) cells, providing a total memory-capacity of 2 Mega ternary words (512 kB). By
activating 16 rows simultaneously in each of these arrays, we can perform 8196 parallel dot product computations with an input vector length of 16. The peripheral circuitry of the SiTe-CiM
I/ SiTe-CiM II array consists of ADCs and small compute elements to sense the MAC outputs and perform partial-sum reduction, respectively. An important distinction between the
SiTe-CiM I and SiTe-CiM II is that we use two ADCs in the former, compared to one in the latter per column, as discussed previously. We compare the SiTe-CiM I/II systems designed with 8T-SRAM, 3T-eDRAM and 3T-FEMFET arrays with near memory baseline architectures corresponding to the memory technologies. For the baselines, we perform the dot product computations and partial-sum reduction in the near-memory compute (NMC) units, the inputs to which are read in a sequential row-by-row manner from each memory array. We design two variants of the near-memory baseline designs – (i) iso-capacity and (ii) iso-area. The iso-capacity baselines contain 32 memory arrays of size 512x256 binary cells (256x256 ternary cells). To design the iso-area baseline architectures, we utilize (a) 41, 48 and 47 8T-SRAM, 3T-eDRAM and 3TFEMFET memory arrays of size 512x256 for comparison with SiTe CiM I and (b) 38, 42 and 41 8T-SRAM, 3T-eDRAM and 3T-FEMFET
memory arrays of size 512x256 for comparison with SiTe CiM II. For the iso-area comparison, the number of SiTe-CiM I and II arrays (=32) are lesser in number than their NM counterparts. This is because SiTe-CiM I and II cell area is larger than their NM counterparts. Moreover, as discussed before,  the area-overhead of the ADCs in SiTe CiM designs outweighs that of the multiply-accumulate unit in the NM baselines, leading to an overall increase in size of the SiTe CiM macros.

\subsection{Performance analysis}

\begin{figure}[t!]
\centering
  \includegraphics[width = 0.98\linewidth]{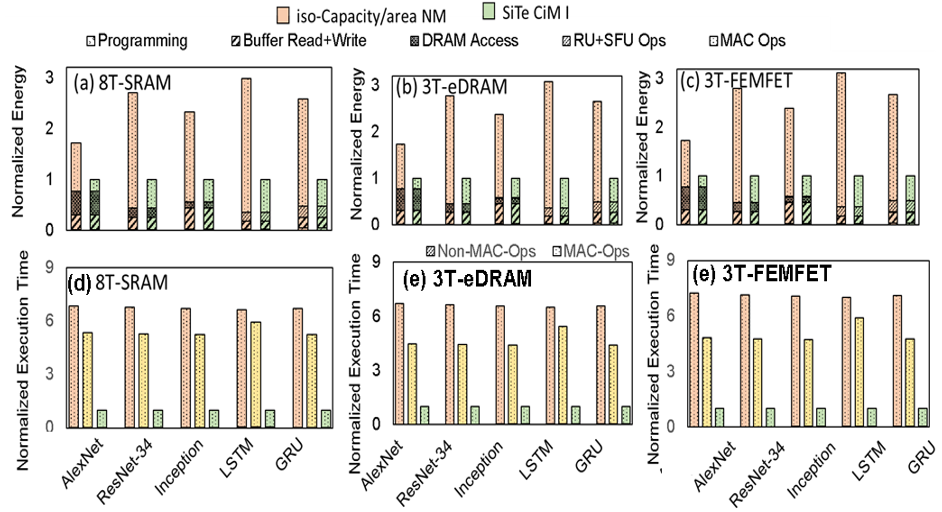}
  \caption{Normalized energy of SiTe CiM I designed with (a) 8T-SRAM, (b) 3T-eDRAM and (c) 3T-FEMFET and normalized execution time of of SiTe CiM I designed with (d) 8T-SRAM, (e) 3T-eDRAM and (ff) 3T-FEMFET with respect to iso-capacity and iso-area baselines using respective near memory baselines for a suite of DNN benchmarks.}
  \label{fig:site1_system}
\end{figure}

We show performance benefits of 8T-SRAM, 3T-eDRAM and 3T-FEMFET based SiTe-CiM I (Fig. ~\ref{fig:site1_system}(a-c)) and SiTe-CiM II (Fig. ~\ref{fig:site2_system}(a-c)) over the corresponding iso-capacity and iso-area NM baselines designed. The performance improvements over the near-memory baselines arise from the massively parallel CiM operations in both SiTe-CiM I and SiTe-CiM II. 

When SiTe CiM I is designed with 8T-SRAM, 3TeDRAM and 3T-FEMFET, we obtain 6.74X, 6.59X and 7.12X average speed-up, respectively over the iso-capacity baselines. Further, we observe 5.41X, 4.63X, 5X average speed-up of 8T-SRAM, 3T-eDRAM and 3T-FEMFET SiTe-CiM I over the iso-area NM baselines respectively, across the benchmarks considered.

Similarly, SiTe CiM II designed with 8TSRAM, 3T-eDRAM and 3T-FEMFET exhibit 4.9X, 4.78X and 5.06X latency reduction compared to their respective iso-capacity baselines. When compared to iso-area NM baseline, the improvements are 4.21X, 3.85X and 3.99X, respectively.

\subsection{Energy benefits}

\begin{figure}[t!]
\centering
  \includegraphics[width = 0.98\linewidth]{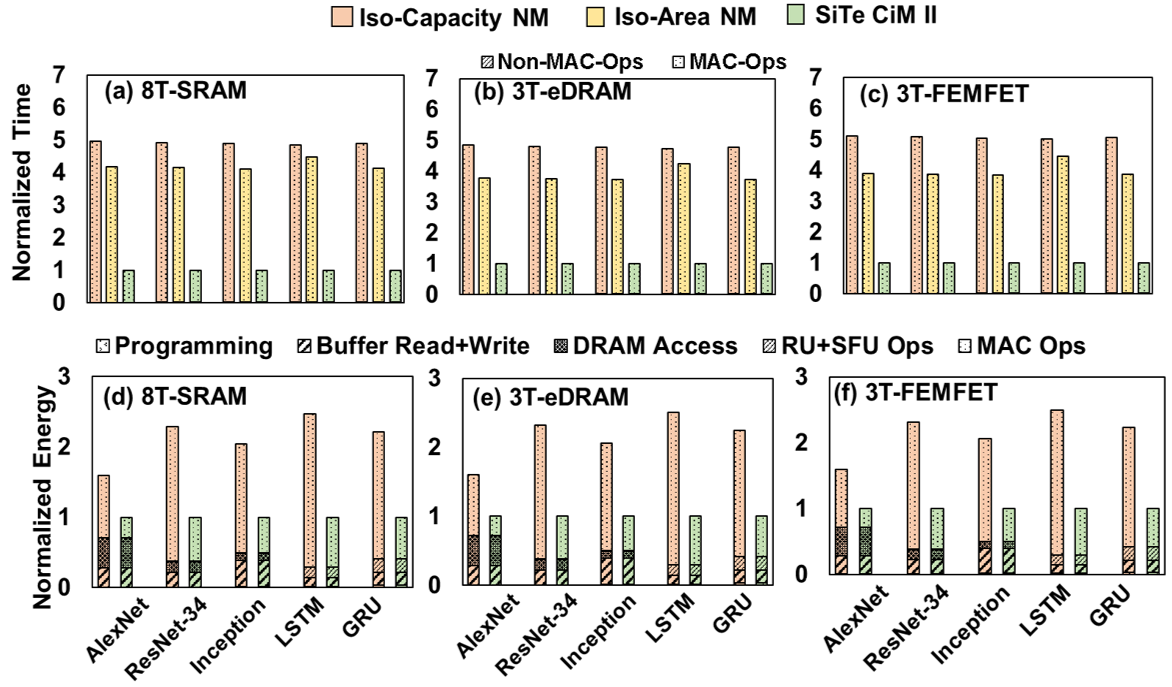}
  \caption{Normalized execution time of SiTe CiM II designed with (a) 8T-SRAM, (b) 3T-eDRAM and (c) 3T-FEMFET and normalized energy of of SiTe CiM I designed with (d) 8T-SRAM, (e) 3T-eDRAM and (f) 3T-FEMFET with respect to iso-capacity and iso-area baselines using respective near memory baselines for a suite of DNN benchmarks.}
  \label{fig:site2_system}
\end{figure}

Let us now discuss the system-level energy benefits of SiTe-CiM I and SiTe-CiM II compared to the iso-area and iso-capacity near memory baselines. The energy benefits over both the baselines are similar since the total energy depends on the total number of 
operations, which remain the same across the two baselines. Fig ~\ref{fig:site1_system} (d-f), ~\ref{fig:site2_system} (d-f) shows the system-level energy benefits of SiTe CiM I and SiTe CiM II. Let us discuss SiTe CiM I first. We achieve 2.46X, 2.52X and 2.54X average energy reduction in 8T-SRAM, 3T-eDRAM and 3T-FEMFET compared to their corresponding NM baselines, respectively, across the benchmarks considered. Similarly, we achieve 2.12X, 2.14X
and 2.14X average energy reduction in 8T-SRAM, 3T-eDRAM and 3T-FEMFET based SiTe CiM II compared to their corresponding NM baselines. Overall, the superior energy efficiency of the proposed SiTe CiM system is due to the parallelism offered by the SiTe-CiM arrays as a result of multi-word line assertion.
\section{Applicaton of SiTe CiM to Other Memory Technologies}
\label{sec:othertech}

In this work, we have shown the applicability of SiTe CiM for three technologies which feature separate read and write paths. However, the proposed technique is not limited to these three technologies and can be extended to other memory technologies as well. Here, we provide a brief perspective on the implications of using SiTe CiM 1 and SiTe CiM II for technologies with shared and separate read-write paths.

Let us begin with memories with separate read and write paths, in which CiM is decoupled from write and can be independently optimized. In general, SiTe CiM 1 can be applied irrespective of the storage mechanism, as long as the underlying bit-cells ($M_1$ and $M_2$) comprise a read access transistor. Without the read access transistor, cross-coupling of bit-cells is not possible. One option to get around this limitation is to add a read access transistor to the bit-cell, albeit at the cost of area. SiTe CiM II is more general and can be applied seamlessly even to memories without a read access transistor (such as 1-FeFET \cite{fefet}), as it uses four shared transistors per sub-column for cross-coupling and does not rely on the read access transistor of an individual bit-cell.

Now, let us consider memories with shared read and write paths, and in particular 1-transistor 1-resistor (1T1R) non-volatile memories. (Note, standard memories with shared read-write paths such as  6T SRAMs and 1T-DRAMs are not suitable for CiM, in general, due to stability concerns \cite{xsram} and destructive read \cite{1tdram}, respectively, and therefore, are not considered in this discussion). Applying SiTe CiM I to 1T-1R to achieve the signed ternary CiM functionality is straightforward and would involve adding two cross-coupling transistors for the interaction of the two 1-T-1R bit-cells composing the ternary weight. However, note, the sizing of the access transistor in the 1T-1R structure is not only dictated by the read/CiM operation, but also by the write operation. Furthermore, the size of the cross-coupling transistors need to match that of the access transistor in the 1T-1R bit-cell for correct CiM functionality (to achieve similar resistance in the CiM path for input = 1 and -1). Thus, the cost of adding the cross-coupling transistors \textit{may be} more for 1T-1R (compared to the memories with separate read and write paths), depending on the write specifications. However, in principle, the signed ternary CiM functionality is possible by utilizing SiTe CiM I design in the 1T-1R NVMs.  On the other hand, it is more challenging to apply SiTe CiM II to 1T-1R memories as adding the four shared transistors in a sub-column would introduce an additional series-connected transistor in the write path, potentially leading to loss in write speed/efficiency and in more dire cases, write failures. One possible way to get around this issue is to increase the sizes of the access transistors in the 1T-1R bit-cell and the four shared transistors, albeit at the cost of area.

\section{Conclusion}
\label{sec:conclusion}

In this work, we propose a technique that enables the design of compute-in-memory (CiM) cells capable of performing vector-matrix multiplications in \textit{signed} ternary (SiTe) regime for deployment in ternary DNN accelerators. We present two flavors of our SiTe CiM designs: (i) aimed at high CiM performance and energy efficiency (SiTe CiM I) and (ii) targeted for high area efficiency (SiTe CiM II).  In the first flavor (SiTe CiM I), our approach  involves cross-coupling of underlying binary bit cells with two additional transistors to enable CiM of dot product of signed ternary inputs and signed ternary weights. The second flavor (SiTe CiM II) reduces the area overhead of SiTe CiM I by cross-coupling sub-columns (rather than individual bit cells).  We establish generality of our technique by implementing it for three different technologies: 8T-SRAM, 3T-eDRAM and 3T-FEMFET, each of which feature a separated read-write path. We show that the proposed designs achieve CiM of dot products in the signed ternary regime with an area penalty of 18\%-34\% for SiTe CiM I and 6\% for SiTe CiM II. SiTe CiM arrays by virtue of the bit-cell/sub-column cross-coupling in conjunction with multi-word line assertion enable massively parallel and efficient computation of dot-products  of signed ternary inputs and weights.  At the array level, up to 88\% higher CiM performance and 78\% lower CiM energy are achieved compared to the near-memory baselines (at the cost of higher area and read/write speed/energy). Efficient dot product computations (the most dominant kernel) translate to significant system-level benefits of SiTe CiM. We investigate the system-level implications by analyzing a suite of DNN benchmarks for convolutional and recurrent neural networks using the proposed SiTe CiM designs and show an average system performance boost of up to 7X and and energy efficiency increase of up to 2.5X (compared to the near-memory designs) .

\section{Acknowledgement}
\label{sec:ack}

The authors thank Chunguang Wang from Purdue University for his input on FEMFET transistor modeling.

\bibliography{references.bib}
\bibliographystyle{IEEEtran}

\end{document}